\documentstyle[12pt,aasms4] {article}

\begin{document}

\title{V(RI)$_{\rm {\boldmath C}}$ Photometry of Cepheids in the Magellanic Clouds}

\author{Thomas G. Barnes III\altaffilmark{1}}
\affil{McDonald Observatory, The University of Texas at Austin, Austin, TX 78712--1083; 
tgb@astro.as.utexas.edu}

\author{Inese I. Ivans, Jodie R. Martin and C. S. Froning}
\affil{Department of Astronomy, The University of Texas, Austin, Texas 78712--1083;
iivans@astro.as.utexas.edu; jodie@astro.as.utexas.edu; cyndi@astro.as.utexas.edu}

\and

\author{Thomas J. Moffett}
\affil{Purdue University, Department of Physics, 1396 Physics Building, West Lafayette, Indiana
47907--1396; moffett@physics.purdue.edu}

\altaffiltext{1}{Visiting Astronomer, University of Toronto Southern Observatory, Las Campanas,
Chile}

\begin{abstract}
We present {\it V(RI)}$_{C}$ data for thirteen Cepheids in the Large Magellanic 
Cloud and eight in the Small Magellanic Cloud. The total number of new measures is 
fifty-five in each wavelength band.  The median uncertainty in the photometry is $\pm 0.03$ mag.
These results supplement a larger photometric program by Moffett, Gieren \& Barnes
\markcite{m1}(1998) which contained 1000 measures ($\pm 0.01$ mag) in each wavelength band on 22
variables with periods in the range 8--133 days.

\noindent
{\it Subject headings:} Magellanic Clouds --- Cepheids --- stars: fundamental parameters
(magnitudes, colors)

\end{abstract}

\section{Introduction}
This is the fourth paper in a program to determine distances to the Large and Small 
Magellanic Clouds using the visual surface brightness technique. This technique 
provides Cepheid distances which are essentially independent of reddening and are independent of 
the PLC relation and its calibration.  The visual surface brightness technique requires radial
velocities and photometric values of high quality.  Fortunately there are velocity curves for
fourteen LMC and eight SMC Cepheids from the work of the CORAVEL group (Imbert et al.\
\markcite{i1}1985, \markcite{i2}1989) and from Caldwell et al.\ \markcite{c2}(1986).  On the
other hand, existing {\it BV(RI)}$_{C}$ photometry of these stars (Caldwell et al.\
\markcite{c2}1986; Caldwell \& Coulson \markcite{c1}1986) samples the light curves too sparsely
for our analysis, hence our present program to improve the available photometry. 

In Paper 1 we demonstrated the surface brightness technique for distance determination on HV 829 in
the SMC using preliminary photometric data (Barnes, Moffett, \& Gieren \markcite{b2}1993). For HV
829 we obtained a distance modulus of $18.9 \pm 0.2$ mag. Because HV 829 may not lie at the
centroid of the SMC, this may not be the mean distance to the SMC. 

In Paper 2 we presented new Cousins {\it BV(RI)}$_{C}$ photometry of fourteen Magellanic Cloud
Cepheids and eight Small Magellanic Cloud Cepheids which yielded light curves of high quality,
consistent with the quality of the radial velocity curves (Moffett, Gieren \& Barnes
\markcite{m1}1998) and sufficient for surface brightness analysis.  

In Paper 3 we used the new photometry and existing radial velocities to determine radii for sixteen
Magellanic Cloud Cepheids and to compare those radii with results for Galactic Cepheids (Gieren,
Moffett \& Barnes \markcite{g1}1999).

In this paper we present additional Cousins {\it V(RI)}$_{C}$ photometric data for all but
one of the stars in Paper 2. The present data were actually the first to be obtained in our CCD
observing program, but because the observing program shifted to another telescope for all
subsequent runs, the present data became `orphaned' and have only now been reduced for
publication.  A follow-up paper will use the full set of photometry to determine individual
Cepheid distances from the visual surface brightness technique.  These data are also useful for
other distance techniques, $\it e. g.$ the infrared flux method.

\section{Data Acquisition}

The observations were acquired at Las Campanas Observatory, Chile, in the period 2 ­- 10 October
1990, using the 61 cm (f/15) Helen Sawyer Hogg Telescope of the University of Toronto Southern
Observatory. (We regret to say that the University of Toronto Southern Observatory no longer
exists.)  The data system was a Photometrics 512 x 512 CCD with 20 micron pixels providing  a 3.9
arc minute field of view.  We observed with Cousins system filters {\it V(RI)}$_{C}$.   

Integration times for the standard stars ranged from 0.4 sec to 12 sec; integration times for the
Cepheids ranged from 10 sec to 900 sec. 

Flat field observations were made on a dome flat every night and on evening twilight sky on two
nights.  

Useful data were obtained on six nights of the nine nights; four nights were photometric and two
were partly cloudy.

\section{Data Reduction}

The CCD frames were reduced using standard IRAF$^{1}$ procedures (Tody \markcite{t1}1993).
\footnotetext{IRAF is distributed by the National Optical Astronomy Observatories, which are
operated by the Association of Universities for Research in Astronomy, Inc., under cooperative
agreement with the National Science Foundation.} Shutter timing corrections were made using an
approximation appropriate to this CCD system kindly provided by Ian Shelton.  The images were
bias--corrected and flat--fielded using the twilight sky flats.  We employed the DAOPHOT (Stetson
\markcite{s1}1987) package within IRAF to obtain instrumental magnitudes in these generally crowded
fields by selecting a number of well--sampled stars in uncluttered regions to compute the point
spread function. In the Cepheid fields, up to twenty comparison stars, chosen from those calibrated
in Paper 2, were also measured.  

Each night extinction pairs from the list of Barnes \& Moffett \markcite{b1} (1979) were observed,
and on the photometric nights standard stars from the list of Landolt \markcite{l1}(1992) were
observed.  Atmospheric extinction corrections and transformation terms were determined within
IRAF.  Mean transformation coefficients were formed and applied to all nights to reduce the data
to the standard Cousins system.

Each Cepheid measurement was adjusted in zero point differentially against the mean of the
comparison stars in its field.  This not only improved the quality of the Cepheid measures but also
permitted observations from the two partly cloudy nights to be reduced to the standard system.  Mean
values for the comparison stars were adopted from the results in Paper 2 and were not re-determined
here.  The number of comparison stars used for an image varied from two to twenty with a median of
ten.  Iterative two--sigma rejection was used to discard outliers in the comparison star measures.
The median number discarded was one comparison star, and the maximum number was six of twenty
comparisons.  

\section{Photometric Results}

Table 1 gives our photometric results. Separate Julian Dates are given for each passband because
of the long integration times in some of the exposures.  

Table 1

The uncertainty $\sigma$ given for each Cepheid is the standard deviation in the comparison star
values on the appropriate image. Because comparison star magnitudes (on the Cousins system) were
adopted from Paper 2, the scatter in their individual measures about the adopted means is a
reasonable representation of the uncertainty in a single stellar magnitude measurement on that
image, incorporating the errors in magnitude determination, atmospheric extinction
and transformation to the standard system.  The Cepheid was almost always the brightest star on the
image making the quoted uncertainties conservative estimates. The median standard deviation in the
comparison star measures is $\pm 0.043$ mag.\ in $\it V$, $\pm 0.033$ mag.\ in $\it R$$_C$, and
$\pm 0.034$ mag.\ in $\it I$$_C$.  

A check on how well the present photometry fits the photometric system of Paper 2 was made by
comparing the values in Table 1 to the light curves in Paper 2. We formed a difference (Table 1
$\it minus$ Paper 2) for all measures in Table 1 which fell within 0.02 in phase to a value in
Paper 2.  Based on thirty-five differences we found $\Delta V$ = 0.006 $\pm 0.050$ mag.\, $\Delta R$
= 0.009 $\pm 0.034$ mag.\, and $\Delta V$ = 0.005 $\pm 0.034$ mag.  The current photometry clearly
match the {\it V(RI)}$_{C}$ photometric system of Paper 2. This is also illustrated in Fig. 1.

Fig. 1

The scatter in these differences is a measure of the combined uncertainties of Table 1 and Paper
2. Removing in quadrature the uncertainty of $\pm 0.01$ mag.\ quoted in Paper 2, the uncertainties
for Table 1 are estimated to be $\pm 0.049$ mag.\ in $\it V$, $\pm 0.032$ mag.\ in $\it R$$_C$, and
$\pm 0.032$ mag.\ in $\it I$$_C$.  These are nearly identical to the uncertainties from the
scatter in comparison star measures on-frame and thus confirm that our adopted uncertainties are
realistic. 

The present results do not achieve the quality of those in Paper 2, which were based on data
using CCD/filter systems on the CTIO 0.9 m telescope.  This may be attributed to the more compressed
plate scale and smaller field size (fewer comparison stars) on the 0.6 m, a less refined image
calibration procedure at the telescope in this first observing run, and fewer nights of data from
which to calibrate the photometric system.  The importance of these measures lies principally in the
phase gaps which they fill in the overall data set.

Fig. 2

The data reported here have been added to the McMaster Cepheid Photometry and Radial Velocity Data
Archive maintained by Doug Welch at URL http://www.physics.mcmaster.ca/Cepheid/. 

\acknowledgments
These observations were made possible by a generous allotment of observing time from the
University of Toronto.  TGB thanks Pablo D. Prado for training on the telescope and CCD
system, Brian Beattie for converting the magnetic tapes to Exabyte tape, and Ian Shelton for 
help with the shutter correction. Financial support is gratefully  acknowledged from NATO Grant
900494 (TGB) and from McDonald Observatory.  The hospitality of the Las Campanas Observatory is
gratefully acknowledged.

\newpage

\figcaption{The {\it V} light curve and {\it (V-R)}$_{C}$ and {\it (V-I)}$_{C}$ color curves for
the SMC Cepheid HV 1338 (P = 8.5 days), the faintest star in our sample. Crosses are used
for photometry from Paper 2 and filled symbols for photometry from this paper (Table 1). Phases are
computed from the arbitrary epoch HJD 2440000 using the period determined in Paper 2.} 

\figcaption{The {\it V} light curve and {\it (V-R)}$_{C}$ and {\it(V-I)}$_{C}$ color curves for
the LMC Cepheid HV 2338 (P = 42.2 days).  The observation at phase 0.62 lies in a crucial phase
interval. Crosses are used for photometry from Paper 2 and filled symbols for photometry from
this paper (Table 1). Phases are computed from the arbitrary epoch HJD 2440000 using the period
determined in Paper 2.}


\begin{references}

\reference{b1}
Barnes, T.G., \& Moffett, T.J.\ 1979, \pasp, 91, 289
\reference{b2}
Barnes, T.G., Moffett, T.J., \& Gieren, W.P.\ 1993, \apj, 405, L51 (Paper 1)
\reference{c1}
Caldwell, J.A.R., \& Coulson, I.M.\ 1986, SAAO Circ., 8, 1
\reference{c2}
Caldwell, J.A.R., Coulson, I.M., Spencer Jones, J.H., Black, C.A., \& Feast, M.W.\ 
1986, \mnras, 220, 671
\reference{g1}
Gieren, W. P., Moffett, T. J., \& Barnes, T. G.\ 1999, \apj, 512, 553 (Paper 3)
\reference{i1}
Imbert, M., Andersen, J., Ardeberg, A., Bardin, C., Benz, W., Lindgren, H., Martin, N., 
Maurice, E., Mayor, M., Nordstr\"{o}m, B., \& Pr\'{e}vot, L.\ 1985, \aaps, 61, 
259
\reference{i2}
Imbert, M., Andersen, J., Ardeberg, A., Duquennoy, A., Lindgren, H., Maurice, E., Mayor, M., 
Mermilliod, J.--C., Nordstr\"{o}m, B., \& Pr\'{e}vot, L.\ 1989,  \aaps, 81, 339
\reference{l1}
Landolt, A.U.\ 1992, \aj, 104, 340
\reference{m1}
Moffett, T.J., Gieren, W.P., \& Barnes, T.G.\ 1998, \apjs, 117,135 (Paper 2)
\reference{s1}
Stetson, P. B. \ 1987, \pasp, 99, 191
\reference{t1}
Tody, D. \ 1993, A. S. P. Conf. Series, 52, 173
\end{references}
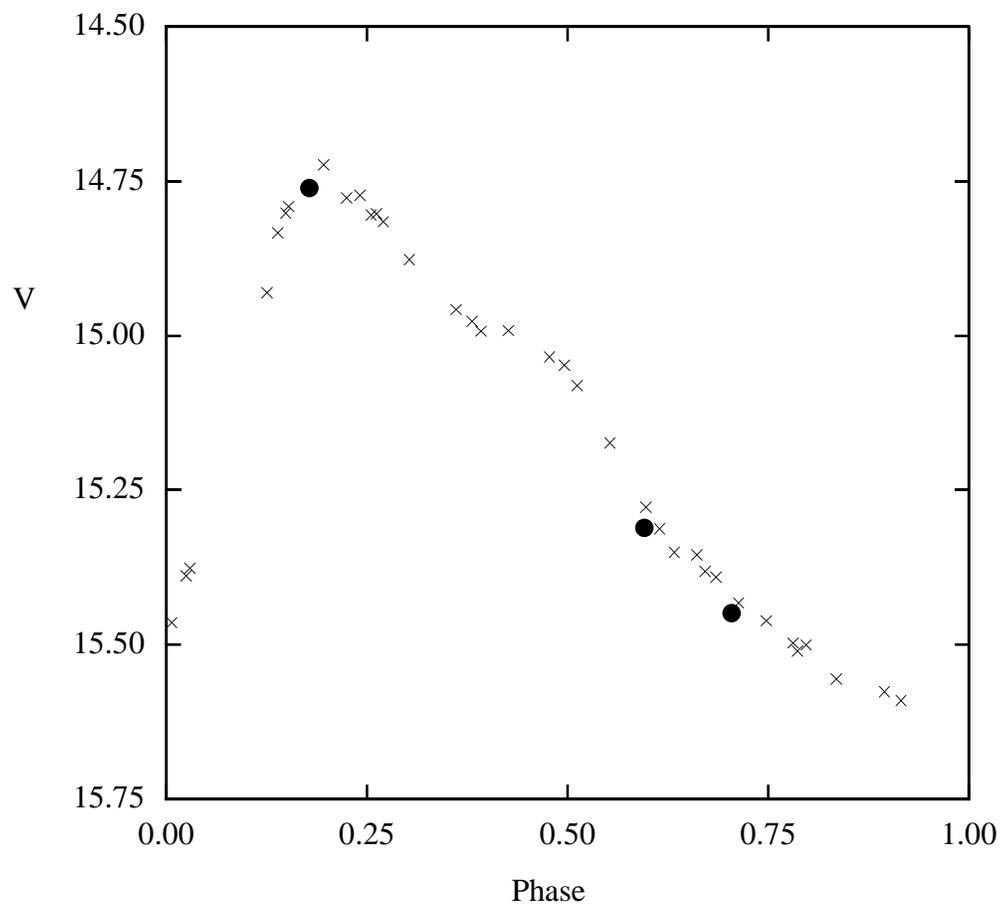
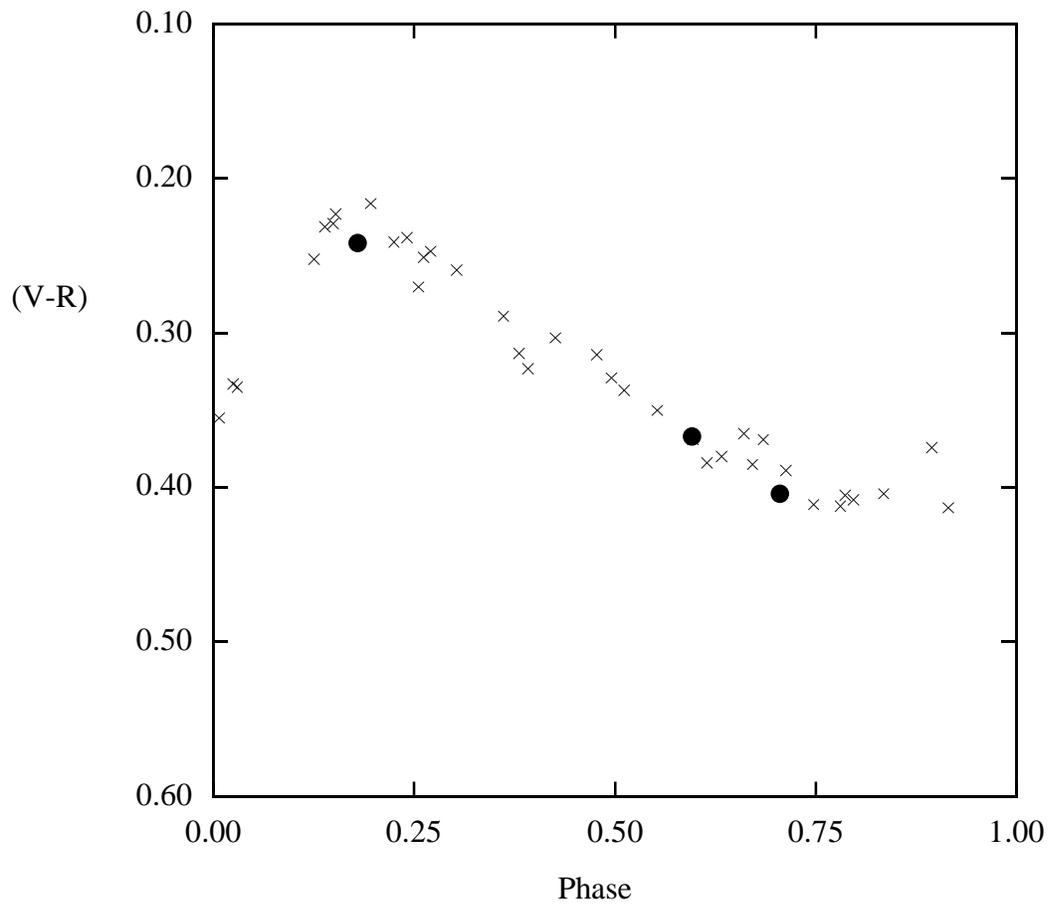
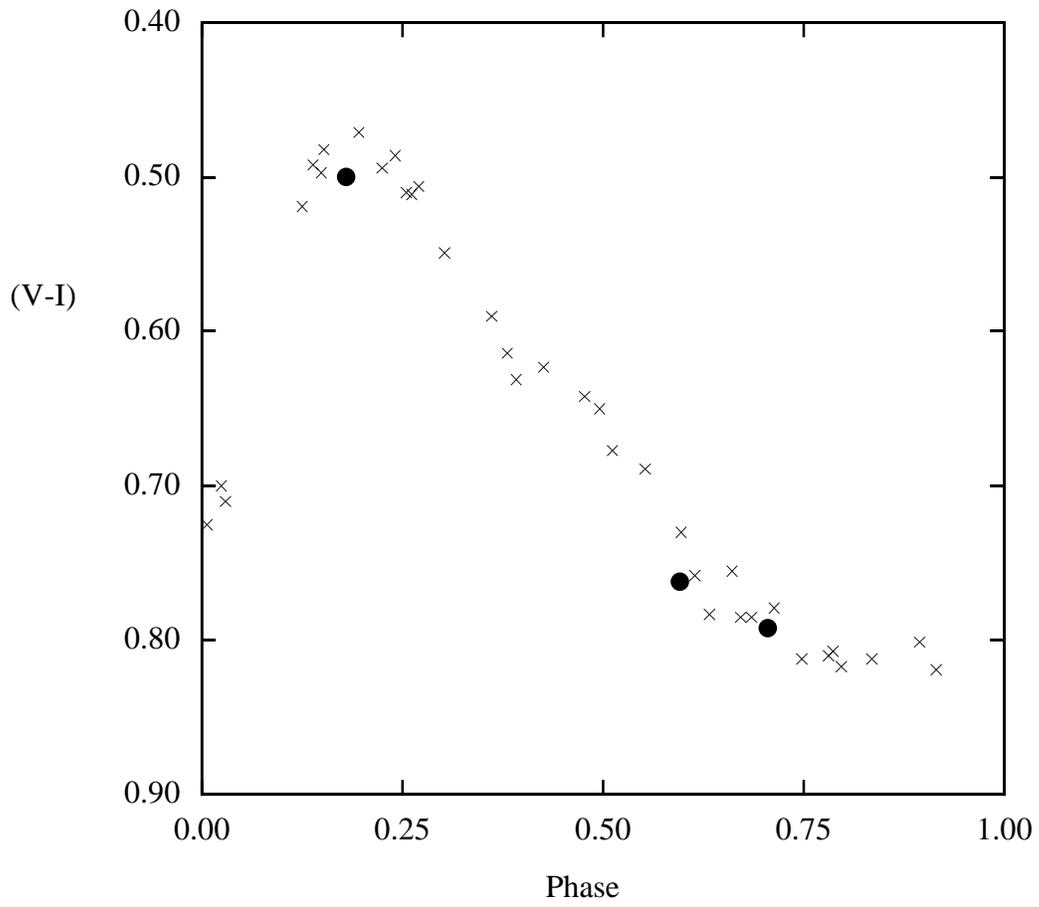
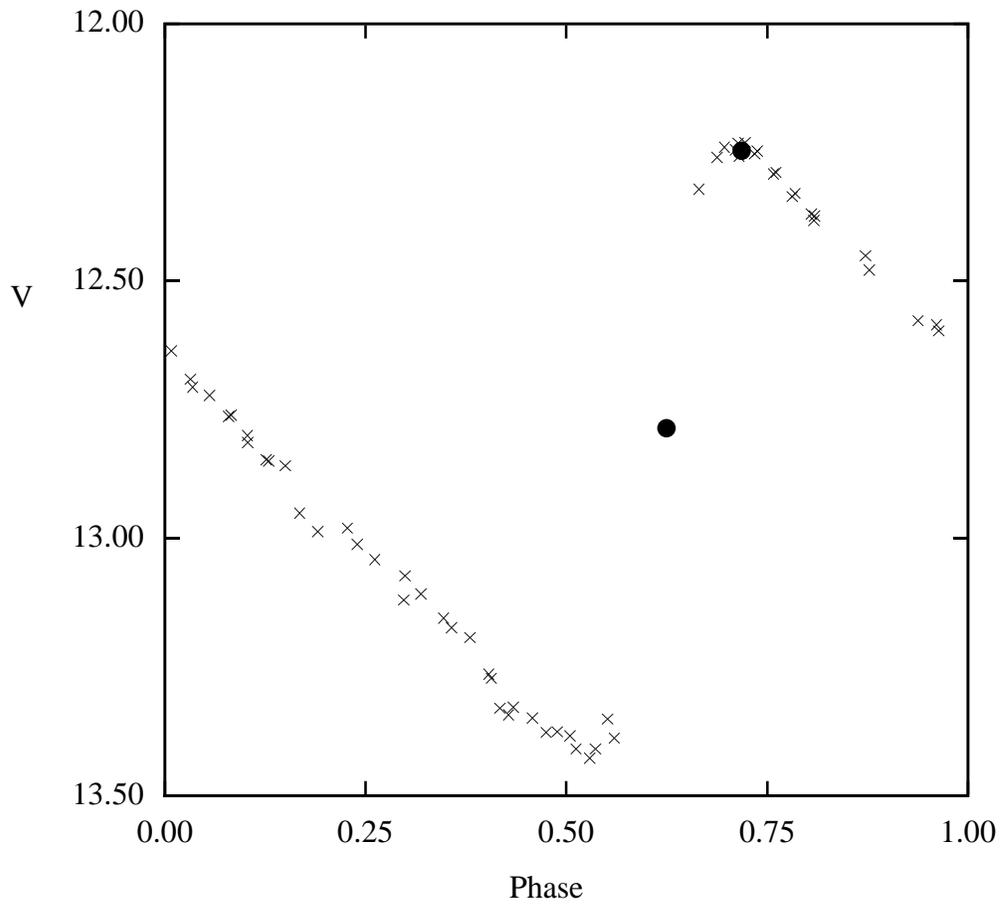
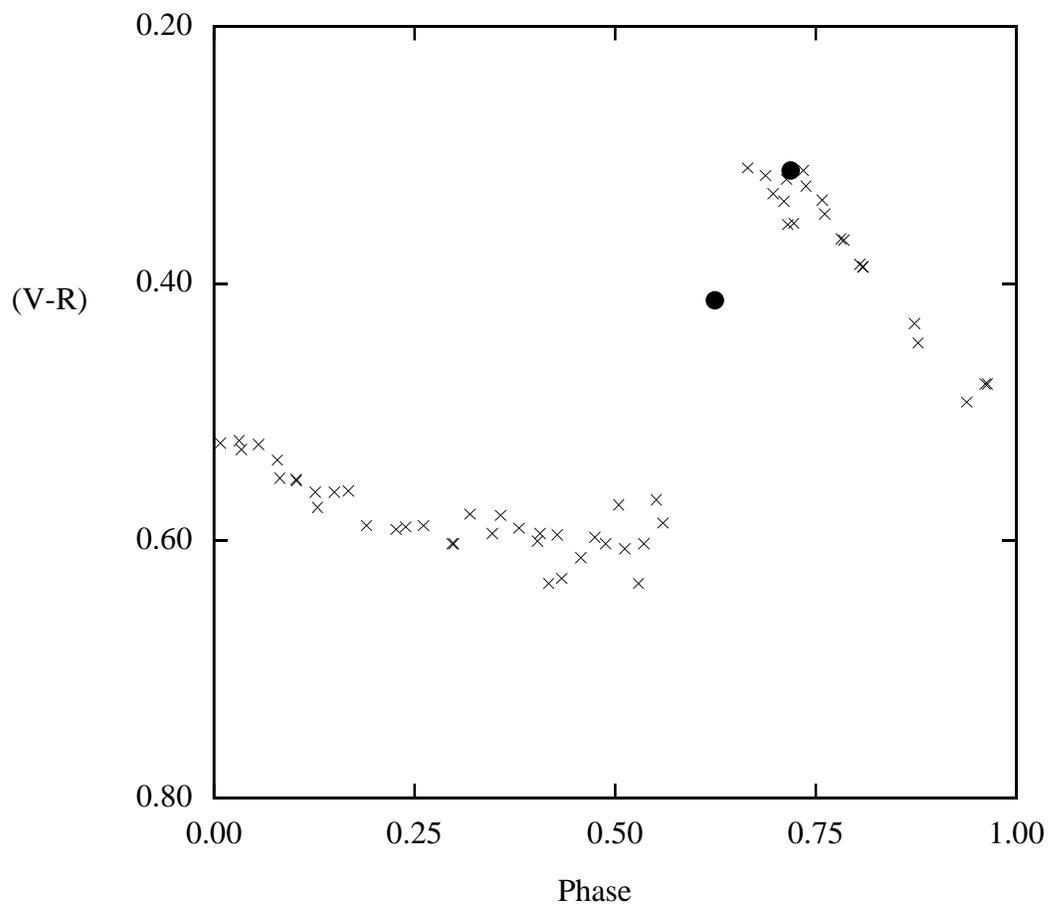
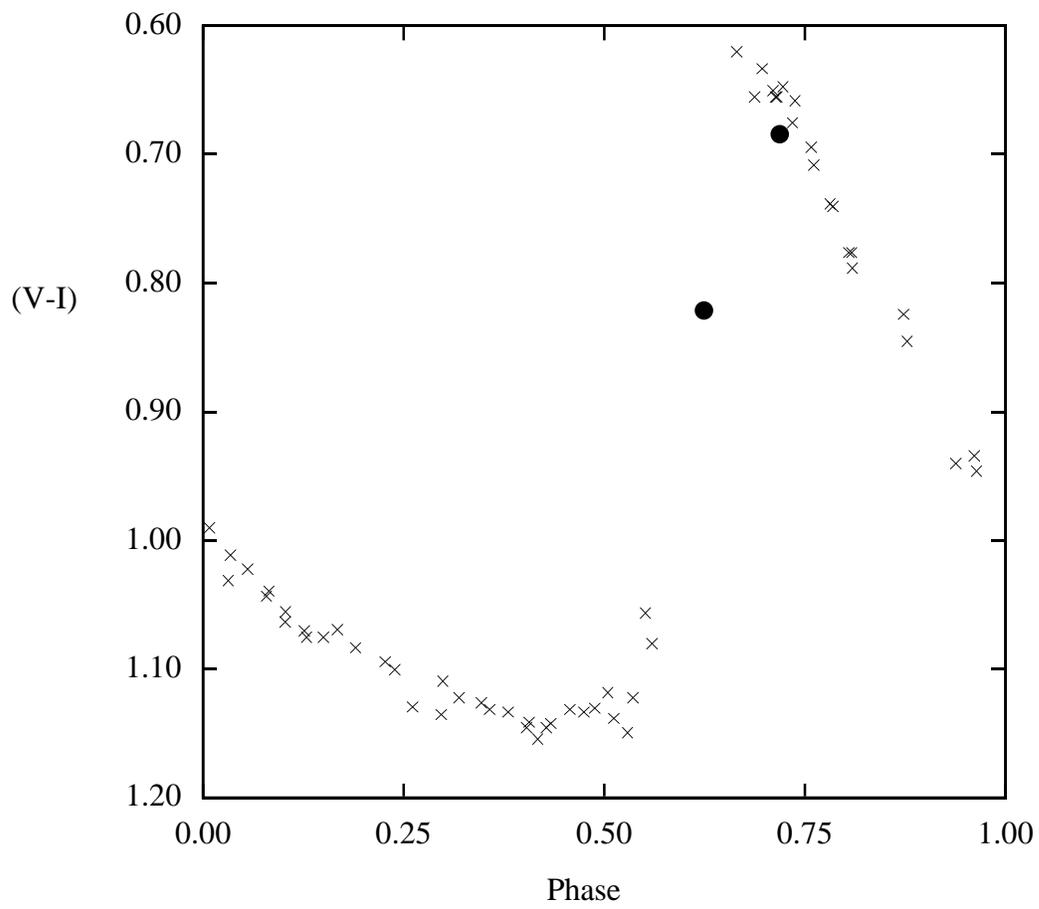
\end{document}